\documentclass[pra,twocolumn,amsmath,amssymb,showpacs,superscriptaddress]{revtex4-1}
\pdfoutput=1 
\usepackage{graphicx}% Include figure files
\usepackage{dcolumn}% Align table columns on decimal point
\usepackage[mathlines]{lineno}
\usepackage{hyperref}
\usepackage{bm}% bold math
\usepackage{epstopdf}
\usepackage{epsfig}
\usepackage{bbold}

\newcommand{\bra}[1]{\ensuremath{\left\langle#1\right|}}
\newcommand{\ket}[1]{\ensuremath{\left|#1\right\rangle}}
\newcommand{\braket}[2]{\ensuremath{\left\langle #1 | #2 \right\rangle}}
\usepackage{color}

\newcommand{\AF}[1]{\textcolor{black}{ #1}}
\newcommand{\IM}[1]{\textcolor{black}{ #1}}
\newcommand{\BV}[1]{\textcolor{black}{ #1}}

\begin{document}

% The following information is for internal review, please remove them for submission
%\widetext
%\leftline{Version xx as of \today}
%\leftline{Primary authors: Joe E. Physics}
%\leftline{To be submitted to (PRL, PRD-RC, PRD, PLB; choose one.)}
%\leftline{Comment to {\tt d0-run2eb-nnn@fnal.gov} by xxx, yyy}
%\centerline{\em D\O\ INTERNAL DOCUMENT -- NOT FOR PUBLIC DISTRIBUTION}
% the following line is for submission, including submission to the arXiv!!
%\hspace{5.2in} \mbox{Fermilab-Pub-04/xxx-E}

\title{Erasing the orbital angular momentum information of a photon}

\author{Isaac~Nape}
\affiliation{School of Physics, University of the Witwatersrand, Private Bag 3, Wits 2050, South Africa}
\author{Bienvenu~Ndagano}
\affiliation{School of Physics, University of the Witwatersrand, Private Bag 3, Wits 2050, South Africa}
\author{Andrew~Forbes}
\email[Corresponding author: ]{andrew.forbes@wits.ac.za}
\affiliation{School of Physics, University of the Witwatersrand, Private Bag 3, Wits 2050, South Africa}
\date{\today}

\begin{abstract}
\noindent Quantum erasers with paths in the form of physical slits have been studied extensively and proven instrumental in probing wave-particle duality in quantum mechanics. Here we replace physical paths (slits) with abstract paths of orbital angular momentum (OAM).  Using spin-orbit hybrid entanglement of photons we show that the OAM content of a photon can be erased with a complimentary \BV{polarization projection} of one of the entangled \BV{pair}.  \BV{The result is the (dis)appearance of azimuthal fringes based on whether the ``which-OAM'' information was erased}.  We extend this concept to a delayed \BV{measurement} scheme and show that the OAM information and fringe visibility are complimentary.
\end{abstract}

\maketitle

%%%%%%%%%%%%%%%%%%%%%%%%%%%%%%%%%%%%%%%%%%%%%%%%%%%%%%%%%%%%%%%%%%%%%%%%%%%%%%%%%%%%%%%%%%%%%%%%%%%%%%%%%%%%%%%%%%%%%%%%%%%%%%%%%%%%%%%%%%%%%%%%%%%%%%%%

\section{Introduction}
Wave-particle duality is a salient feature of quantum mechanics and has primarily been observed through modern variations of Thomas Young's double slit experiments \cite{ rauch1974test, grangier1986experimental, zeilinger1988single,taylor1909interference,gerlich2011quantum}. When the paths \IM{of the double slit} are indistinguishable, multi-path interference results in fringes of high visibility ($V$) in the far-field, which is a characteristic trait of wave-like behavior. Conversely, if the paths are distinguishable ($D$), for example, through the use of which-path markers, the fringes disappear \BV{(particle behavior)}. The physical implications of this are embodied through the principle of complementarity \cite{bohr1928quantum}, emphasizing the mutual exclusivity that exists between complimentary observables. The special case of $D=0$ and $V=1$  corresponds to a maximal observation of interference fringes while that of $D=1$ and $V=0$ corresponds to a full obtainment of the which-path information. Intriguingly, it is permitted to have partial visibility and partial distinguishability, where the result cannot be explained exclusively by a wave-like or particle-like interaction \cite{wootters1979complementarity, greenberger1988simultaneous, jaeger1995two,englert1996fringe}, and this may quantitatively be expressed \BV{through the following inequality:} $V^{2}+D^{2}\leq1$. \BV{Thus, gaining knowledge of path information} ($D \neq 0$), reduces the visibility of the fringes ($V < 1$).  Interestingly, the path information can be erased with a complimentary projection with respect to the path markers of the double slit, reviving the interference fringes. 

Scully and co-authors \cite{scully1982quantum,scully1991quantum} proposed such a device, the quantum eraser, which is now ubiquitous in experimental verifications of the complementarity principle. For example, in the double slit experiment presented in \cite{walborn2002double}, a polariser is used to recover the interference pattern that is lost due to path distinguishability with circular polarisers. By orienting a polariser in a diagonal position, the path information is erased. Numerous other experiments have been performed with photonic systems using double slits \cite{neves2009control,neves2009hybrid}, interferometers \cite{kwiat1992observation,herzog1995complementarity, kim2000delayed, ma2013quantum,chen2014revisiting}, and in \BV{delayed-choice} measurement schemes \cite{ma2016delayed, ma2013quantum, jacques2008delayed}. All these experiments have used physical paths to study the multi-path interference in the context of quantum erasers.

Here we generalise the concept of ``path'', showing that it need not be a physical path in the sense of a route through space but an abstract ``path'' in any degree of freedom.  We employ orbital angular momentum (OAM) as our ``path'' and use polarisation as the ``which-path'' marker.  To test this we create hybrid entanglement between photons carrying spin and orbital angular momentum and show control of the fringe visibility through a generalised quantum erasure experiment: the OAM paths marked with polarization do not lead to interference, while introducing the eraser (polarizer) which projects the polarization of one of the entangled photons onto a complementary polarisation basis results in azimuthal fringes with high visibility.  We perform this experiment in both the conventional quantum erasure and \BV{delayed-choice} schemes, in both cases showing \IM{control of} \BV{ the nature of the photons, from particle (no visibility) to wave (full visibility)}.  Our experimental results are in very good agreement with theory, offering a simple approach to illustrate the concept of path in quantum mechanics.

\section{Theory}

\subsection{Revisiting the double-slit quantum eraser}

\begin{figure*}[t]
	\includegraphics[width=\linewidth]{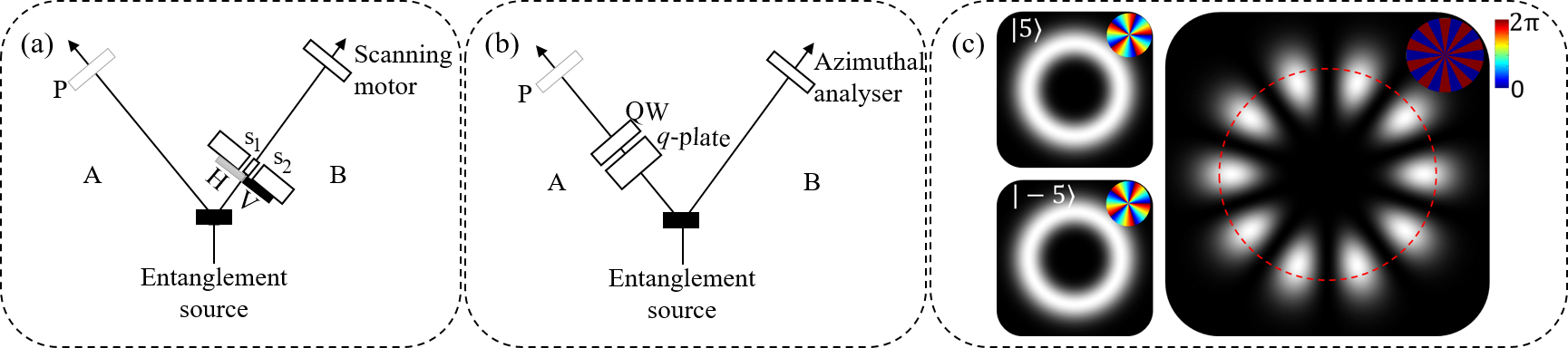}%width=1\linewidth
	\caption{(a) Schematic of a quantum eraser that uses polarization entangled photons using a physical double slit as reported by Walborn \BV{\textit{et al.}} \cite{walborn2002double} is shown. The two slits ($s_1$, $s_2$) are marked with orthogonal horizontal (H) and vertical (V) linear polarizers to distinguish the two paths. The polarizer (P) in arm A acts as the eraser. (b) The proposed quantum eraser using geometric phase control to perform OAM-polarization conversion. The polarization control (P) of photon A sets the OAM interference of photon B.}  
	\label{fig:fig1}
\end{figure*}

It is instructive to revisit the concepts of the traditional quantum erasure experiment, illustrated in Fig.~\ref{fig:fig1} (a),  which we do here briefly for the benefit of the reader.  Consider a photon \BV{traversing} two unmarked slits, which can be represented by the following coherent superposition

\begin{equation}
\BV{\ket{\Phi}=\frac{1}{\sqrt{2}}\left(\ket{\psi_{1}}+\ket{\psi_{2}}\right),}
\end{equation}
\BV{ where $\ket{\psi_{1}}$ and $\ket{\psi_{2}}$  are the non-orthogonal states upon traversing slit $1$ and $2$ (path 1 and path 2), respectively. The spatial probability distribution of the photon after the slits is given by $|\braket{\Phi}{\Phi}|^{2}$ where the interference pattern, a sign of the photon travelling through indistinguishable paths, emerges due to the cross terms $\braket{\psi_i}{\psi_{j}}$ for $i \neq j$. However, the fringes disappear when the paths are marked (distinguishable)\AF{, for example,} with orthogonal polarizers which, we assume are oriented along the horizontal (H) and vertical axis (V)}
\begin{equation}
\BV{\ket{\Phi}=\frac{1}{\sqrt{2}}\left(\ket{\psi_{1}}\ket{H}+\ket{\psi_{2}}\ket{V}\right).}
\label{eq:gle state}
\end{equation}
\BV{Now, the cross terms vanish and $|\braket{\Phi}{\Phi}|^{2} = \sum_{i} |\braket{\psi_i}{\psi_{i}}|^2/2$. Equation \ref{eq:gle state} represent a general state of entangled spatial and polarization degrees of freedom of a single photon. An identical representation can be extended to a two-photon case using entanglement.} To illustrate this, consider the schematic for a system that produces polarization entangled photons given by the following state,
 
 \begin{equation}
 \ket{\Phi}_{AB}=\frac{1}{\sqrt{2}}\big( \ket{H}_{A}\ket{V}_{B}+\ket{V}_{A}\ket{H}_{B} \big),
 \end{equation}
where the subscripts $A$ and $B$ label the entangled photons.
%where $\ket{ H }$ and $\ket{ V }$ represent the orthogonal horizontal and vertical polarization states for photons in arm A and B. 
Inserting a double slit in the path of photon B, with each slit ($s_{1}$ and $s_{2}$) marked with orthogonal linear polarizers yields  

 \begin{equation}
	\ket{\Phi'}_{AB}=\frac{1}{\sqrt{2}}\big( \ket{H}_{A}\ket{s_{1}}_{B} +\ket{V}_{A}\ket{s_{2}}_{B} \big). \label{eq:slitEraser}
	\end{equation}

Equation (\ref{eq:slitEraser}) is a hybrid entangled state where the polarization of photon A is entangled with the slit (path) traversed by photon B. For example, measuring the state $\ket{H}$ for photon A means that photon B traverses through slit $s_{1}$, hence no interference fringes will be observed in the far-field of the double slit since there is path information in the system. However if photon A is projected onto the complimentary diagonal basis, \{$\ket{D},\ket{A}$\}, where \BV{$\ket{D}=(\ket{H}+\ket{V})/\sqrt{2}$ and $\ket{A}=(\ket{H}-\ket{V})/\sqrt{2}$ are the diagonal and anti-diagonal states respectively}, then the following projections hold 

\begin{eqnarray}
&\ket{\Phi'}_{AB}\xrightarrow{\hat{D}_{A}}\frac{1}{\sqrt{2}}\Big(\ket{D}_{A}\big(\ket{s_{1}}_{B} +\ket{s_{2}}_{B}\big)\Big),\label{eq:slitComplimentary1}\\
&\ket{\Phi'}_{AB}\xrightarrow{\hat{A}_{A}}\frac{1}{\sqrt{2}}\Big(\ket{A}_{A}\big(\ket{s_{1}}_{B} -\ket{s_{2}}_{B}\big)\Big)\label{eq:slitComplimentary2},
\end{eqnarray}

where $\hat{D}_{A}$ and $\hat{A}_{A}$ are projection operators associated with the states $\ket{D}$ and $\ket{A}$, \BV{acting on} photon A. Thus the  projections of photon A onto complimentary polarization states collapses photon B into a coherent superposition \BV{of the two paths, consequently recovering} the interference pattern. This means that the which-way path information of photon B has been erased.  
 
\subsection{OAM based quantum eraser}
Now \BV{we exchange the notion of path or slit, for that of orbital angular momentum (OAM).}
%consider the OAM degree of freedom (DoF) of photons.
Photons carrying OAM \cite{allen1992orbital,allen1999iv} have attracted great interest in both classical and quantum studies \cite{franke2008advances, molina2007twisted, rubinsztein2016roadmap}. \BV{OAM modes possess} a transverse spatial distribution characterized by an azimuthal phase of $e^{i\ell\phi}$ such that each photon has an angular momentum of $\pm\ell\hbar$ where the integer $\ell$ represents the twist or helicity of the phase profile.  Since OAM \BV{states} of differing $\ell$ are orthogonal, entanglement may be expressed in this basis where each \BV{photon} OAM subspace is spanned by $\mathcal{H}_{2}=\{\ket{\ell}, \ket{-\ell}\}$. The \BV{detected} distribution (intensity distribution in classical light) of the photons is symmetric and uniform in the azimuth for both basis states, each with an azimuthal helicity in phase of opposite sign.  These properties allow OAM \BV{mode of opposite helicity} to be treated as two paths, \BV{indistinguishable in the intensity domain,} so that one may conceive an OAM quantum eraser as depicted in Fig.~\ref{fig:fig1}(b).  

To create an analogous quantum eraser for OAM we require a hybrid entangled state of OAM and polarisation.  To generate the hybrid entanglement, we consider type I spontaneous parametric down-conversion (SPDC) as a source of entangled photons and employ geometric phase control of one of the entangled pairs using Pancharatnam-Berry phase to execute spin-orbit coupling.

The quantum state of the photon pair produced from a type I SPDC process is

\begin{eqnarray}
\ket{\Psi}=\sum_{\ell} c_{|\ell|}\ket{\ell}_{A}\ket{-\ell}_{B}\ket{H}_{A}\ket{H}_{B},
\label{eq:spdc1}		 
\end{eqnarray}

\noindent where $|c_{\ell}|^{2}$ is the probability of finding photon $A$ and $B$ in the state $\ket{\pm\ell}$.   

The hybrid entanglement between photon A and B is obtained by using geometric phase control to perform an orbit-to-spin conversion in arm $A$ \cite{nagali2010generation, karimi2010spin}. This may be done by using a $q$-plate \cite{marrucci2006optical, marrucci2011spin}, a wave-plate with a locally varying birefringence, that couples the polarization and OAM DoF of light according to \BV{the following rules:}  

\begin{eqnarray}
\ket{\ell}\ket{R}\xrightarrow{q\text{-plate}} \ket{\ell + 2q}\ket{ L}, \label{eq:Qplate1}\\
\ket{\ell}\ket{L}\xrightarrow{q\text{-plate}} \ket{\ell - 2q}\ket{ R }.
\label{eq:Qplate2}
\end{eqnarray}

Here $\ket{ L }$ and $\ket{R}$ represent the left and right circular polarization states and $q$ is the charge of the $q$-plate. The $q$-plate inverts the circular polarization of a photon and imparts an OAM variation of $\pm2q$ depending on the handedness of the input polarization.  Noting that \BV{$\ket{H}=(\ket{R}+\ket{L})/\sqrt{2}$} and applying the transformation of the $q$-plate to photon A transforms Eq. (\ref{eq:spdc1}) to

\begin{align}
	\ket{\Psi}&\xrightarrow{\hat{Q}_{A}}\sum_{\ell}c_{|\ell|}\big(\ket{\ell+2q}_{A}\ket{R}_{A} \nonumber \\ 
	&\qquad+\ket{\ell-2q}_{A}\ket{L}_{A}\big) \ket{-\ell}_{B}\ket{H}_{B},\label{eq:spdc2}	 
\end{align}  

where $\hat{Q}_{A}$ is the transformation of the $q$-plate. Coupling photon A into a single mode fiber imposes the condition $\ell=\pm2q$ on the entangled pair (since the OAM of A and B must now be zero). Subsequently, \BV{the post-selected two-photon state reduces to}
%the superposition in Eq. (\ref{eq:spdc2}) collapses to

\begin{align}
	\ket{\Psi'}_{AB}=\frac{1}{\sqrt{2}}\big(\ket{R}_{A}\ket{\ell}_{B}+\ket{L}_{A}\ket{-\ell}_{B} \big) \label{eq:hybrid1},
\end{align} 

where $\ell=2q$. Equation (\ref{eq:hybrid1}) represents a maximally entangled Bell state where the polarization DoF of photon A is entangled with the OAM degree of freedom of photon B, as desired.\\
To obtain the OAM information of photon B, the circular polarization of photon A is converted to linear polarization using a $\lambda/4$ wave plate inserted after the $q$-plate and \BV{oriented} at $\pi/4$ with respect to the \BV{horizontal}. Therefore Eq. (\ref{eq:hybrid1}) becomes

\begin{align}
	\ket{\psi}_{AB}=\frac{1}{\sqrt{2}}\big(\ket{H}_{A}\ket{\ell}_{B}+e^{i\delta}\ket{V}_{A}\ket{-\ell}_{B} \big)\label{eq:hybrid2}.
\end{align} 

Here $\delta = \pi /2$ is a relative phase after the transformation of the $\lambda/4$ wave-plate.  We note that the OAM ``path'' is marked by polarisation. When one path is selected in this way, no interference appears.  However, just as in the double slit case, a projection of the polarization of photon A onto a complimentary basis state \BV{(diagonal or anti-diagonal) will collapse the state of photon B into a superposition of OAM, $\ket{+\ell} + i\ket{-\ell}$, leading to the emergence of azimuthal intensity fringes} with \BV{angular frequency proportional to $2|\ell|$. An} example for the $|\ell|=5$ subspace is shown in Fig.~\ref{fig:fig1} (c). In this case the OAM ``path'' information is erased.

\subsection{Detection scheme}
Suppose the state of the hybrid entangled system is represented by Eq. (\ref{eq:hybrid2}).  A polarizer orientated at an angle $\alpha$ (with respect to the \BV{horizontal}) in arm A will project photon A onto the following target state 

\begin{equation}
\ket{\alpha}_{A}=\cos(\alpha)\ket{H}_{A}+\sin(\alpha)\ket{V}_{A}
\label{eq:alpha},
\end{equation}

thus allowing the ``path'' to evolve from marked to unmarked by a judicious choice of $\alpha$.  Next, the visibility of fringes in arm B needs to be detected, which may easily be done with scanning detectors (or more expensive camera-based systems).  We instead make use of scanning holograms and a fixed detector as our pattern sensitive detector \cite{forbes2016creation}. We create sector states from superpositions of OAM with a relative intermodal phase of $\theta$:

\begin{equation}
\ket{\theta}_{B}=\big(\ket{\ell}_{B}+e^{i2\theta}\ket{-\ell}_{B}\big).
\label{eq:theta}
\end{equation}

The phase structure of $\ket{\theta}_B$ is \BV{azimuthally periodic,} and allows a measurement of the path (OAM) interference in arm B, analogous to detecting OAM entanglement with Bell-like measurements \cite{leach2009violation, dada2011experimental, fickler2012quantum, mclaren2012entangled}. Thus the fringe pattern (or lack thereof) can be detected by scanning through $\theta$. 

The normalized probability of detection given the two projections is 

\begin{align}
	P(\alpha, \theta)&\propto|\bra{\theta}_{B}\bra{\alpha}_{A}\ket{\psi}_{AB}|^{2} \nonumber\\
	&=\frac{1}{2}(1+\sin(2\alpha)\cos(2\theta+\delta)).
	\label{eq:prob}
\end{align}

$P(\alpha, \theta)$ is synonymous to the coincidence counts of the entangled pair.  When the polarizer is orientated at $\alpha=0$, which corresponds to the $\ket{H}$ polarization state, the probability distribution with respect to $\theta$ is a constant since the path is marked.  Conversely, for $\alpha =\pm\pi/4$ which corresponds to \BV{complimentary polarization projections on $\ket{D}$ or $\ket{A}$,} then  $P(\alpha=\pm\pi/4, \theta)\propto1\pm\cos(\theta+\delta)$ and hence the oscillation is an indication of an interference pattern emerging from a superposition of the OAM paths of photon B. Therefore the which-path (OAM) information has been erased. The fringe visibility is given by
 	
\begin{equation}
V=\frac{P_{min}+P_{max}}{P_{max}+P_{min}}.
\end{equation}

\BV{Here, $P_{min}$ and $P_{max}$ are the maximum and minimum photon probabilities obtained from the \BV{azimuthal} scanning.} The theoretical visibility of the interference fringes with respect to the angle of the polarizer ($\alpha$) is  $V=|\sin(2 \alpha)|$. 

\section{Experimental set-up}

\begin{figure*}[t]
	\centering
	\includegraphics[width=\linewidth]{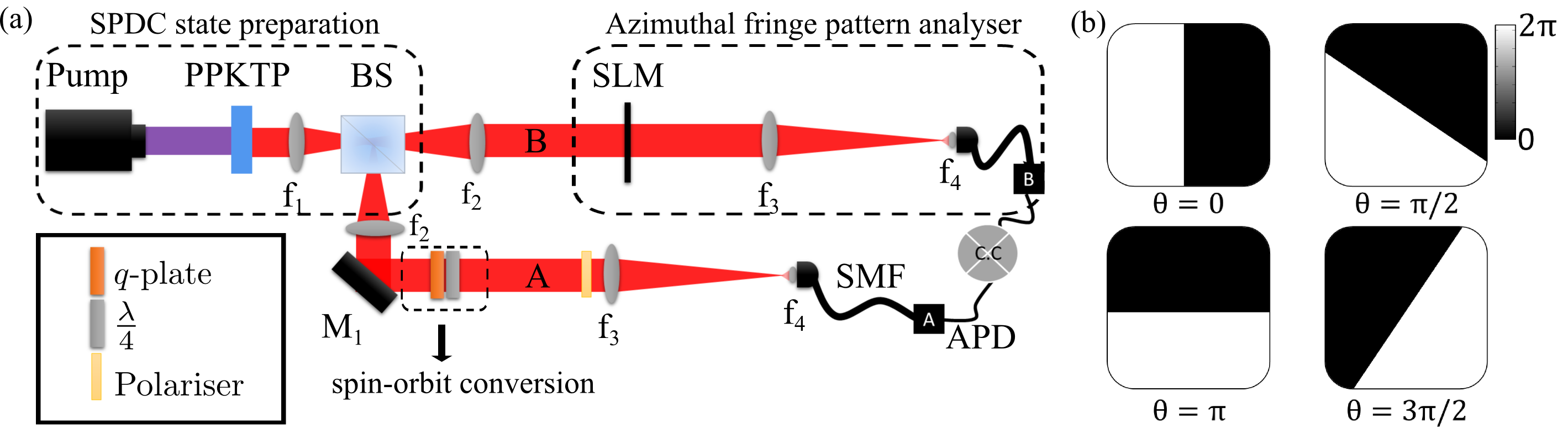}
	\caption{\textbf{(a)} Experimental setup for the hybrid entanglement based quantum eraser. The SPDC state was prepared at the plane of the non-linear crystal (PPKTP) and imaged onto the spatial light modulator (SLM) in arm B. The same imaging system was replicated in arm A where OAM to polarization conversion was obtained through the geometric phase control of photon A using a $q$-plate ($q$=0.5). The imaging system consisted of lenses f$_{1} = 100 $ mm and f$_{2} = 300 $ mm, and by lenses f$_{3} = 500$  mm and f$_{4} = 2$ mm. The photons passed through 10 nm bandwidth interference filters (IF) prior to collection into single mode fiber (SMF). The SMFs were coupled to avalanche photon diodes (APD).   Furthermore, we performed a delay measure type eraser by extending arm A by 2.3 m, corresponding to a relative delay time of 7.66 ns. \textbf{(b)} Angular phase masks that were encoded on the SLM and rotated by an angle $\theta$, serving as an azimuthal scanner to detect azimuthal fringes.   
		}		
	\label{fig:fig2}
\end{figure*}
%%%%%%%%%%%%%%%%%%%%%%%%%%%%%%%%%%%%%%%%%%%%%%%%%%%%%%%%%%%%%%%%%%%%%%%%%%%%%%%%%%%%%%%%%%%%%%%%%%%%%%%%%%%%%%%%%

In Fig.~\ref{fig:fig2} (a), we present the experimental set-up for our quantum eraser with polarization-OAM hybrid entangled photons. A periodically poled potassium titanyl phosphate (PPKTP) nonlinear crystal, cut for type 1  phase matching, was pumped with a 100 mW Coherent Cube  diode laser with a 450 nm nominal  wavelength, producing \BV{collinear} entangled photon pairs at a wavelength of 810 nm.  Each \BV{photon pair was spatially separated in two arms using a $50/50$ beam splitter (BS)}. The spin-orbit conversion was achieved by inserting a $q$-plate \BV{with $q=0.5$ in arm A,} creating polarisation-OAM hybrid entanglement in the $|\ell|= \pm 1$ subspace. In this arrangement, the state of the system is given by Eq.~(\ref{eq:hybrid1}). To mark the states, a $\lambda/4$ wave plate \BV{with fast axis at $\pi/4$ with respect to the horizontal direction, as well as a linear polarizer (eraser), were inserted in arm A. The detection in arm B was performed with binary phase masks shown in Fig.\ref{fig:fig2} (b), encoded on a phase-only spatial light modulator (Holoeye PLUTO) to scan the spatial distribution of photon B; this was done for $\alpha = [0,\pi/4]$ while scanning holograms through $\theta = [0,2\pi]$. The modulated photons were collected with a single mode fiber and measured in coincidence with an integration time of 5 sec.  A gating time of 25 ns was used as a window within which coincidence events recorded with avalanche photo-diodes (Perkin-Elmer) inserted into arms A and B.}
%%%%%%%%%%%%%%%%%%%%%%%%%%%%%%%%%%%%%%%%%%%%%%%%%%%%%%%%%%%%%%%%%%%%%%%%%%%%%%%%%%%%%%%%%%%%%%%%%%%%%%%%%%%%%%%%%%
\begin{figure}[t]
	\centering
	\includegraphics[width=1\linewidth]{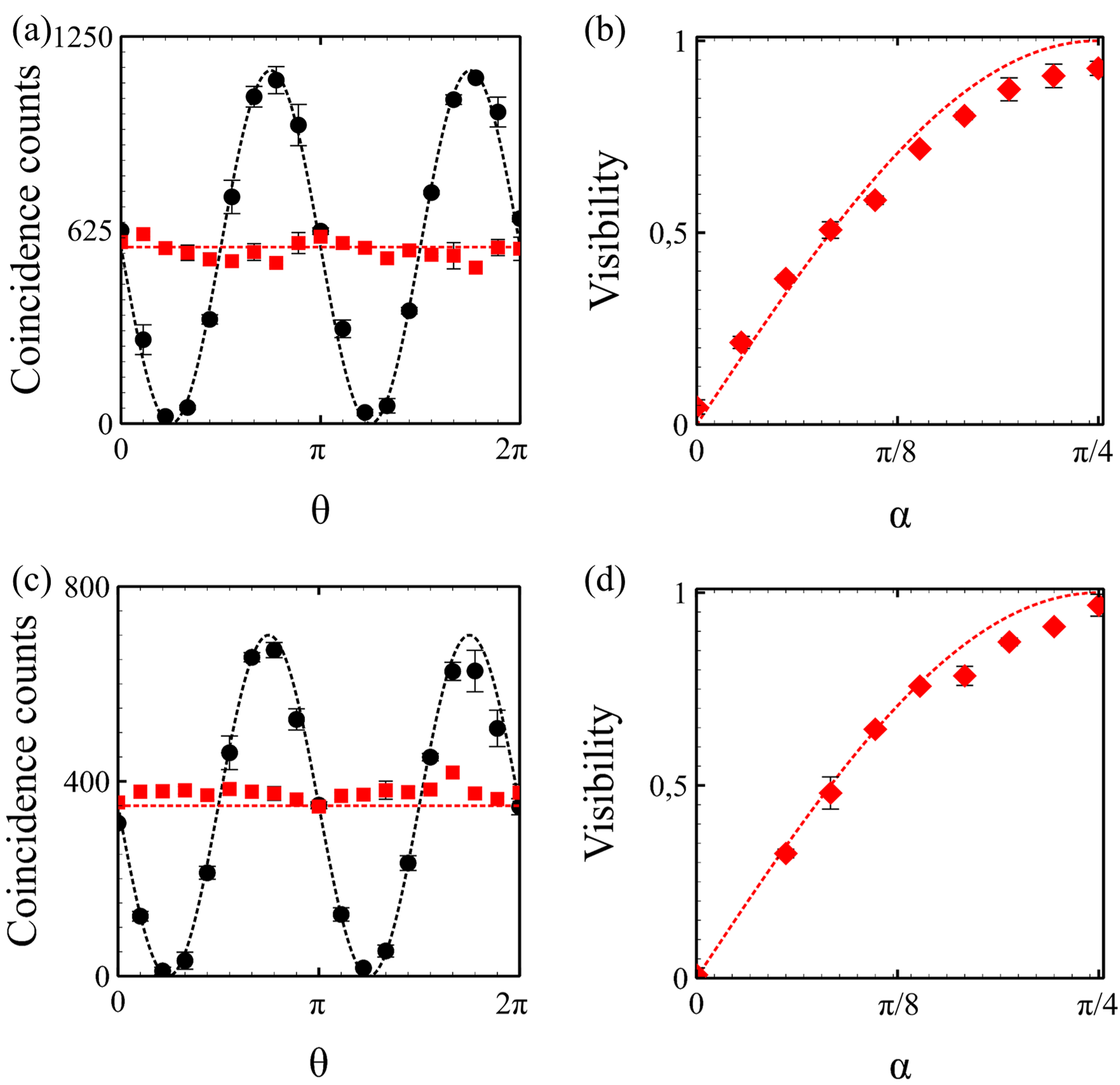}
	\caption{Comparison of theory and experiment for a OAM quantum eraser.  \textbf{(a)} Interference measurements where the OAM path information of photon B is distinguished (squares) and erased (circles) by marking the path with a polarisation choice on photon A. \textbf{(b)} Visibility of interference fringes with a variation of the polariser angle ($\alpha$) in the range 0 to $\pi/4$. Similarly, \textbf{(c)} \BV{delayed-choice} measurement results where the OAM path is distinguished (squares) and erased (circles), and subsequence measurements of the fringe visibility \textbf{(d)} with the polariser angle.  In all panels experimental data is shown as symbols and theoretical calculations as dashed curves.
}	
	\label{fig:fig3}
\end{figure}

\section{Results}
%%%%%%%%%%%%%%%%%%%%%%%%%%%%%%%%%%%%%%%%%%%%%%%%%%%%%%%%%%%%%%%%%%%%%%%%%%%%%%%%%%%%%%%%%%%%%%%%%%%%%%%%%%%%%%%%%%%%%%%%%%%%%%%%%%%%%%%%%%
\BV{The OAM path information of photon B was obtained by projecting photon A onto the states $\ket{H}$ or $\ket{V}$. Here we chose $\ket{H}$, by setting the polarizer in arm A to $\alpha=0$, collapsing the state of photon B to the OAM $\ket{\ell = 1}$.}  The results are presented in Fig.~\ref{fig:fig3} (a), confirming that no interference fringes were observed.  The small oscillations are due to imperfections in the polarization filtering of photon A. The calculated visibility of the interference fringes is $0.04\pm0.01$, in good agreement with the theoretical value of  0.  One can interpret this as photon B carrying a well defined amount OAM, or equivalently, that the OAM path is marked (distinguished) and thus visibility is zero.  The OAM path information was erased by performing a complimentary measurement of the polarization of photon A.  We set the polarizer angle to \BV{$\alpha=-\pi/4$, thus selecting the polarization state $\ket{A}$, collapsing the state of photon B into a superposition of OAM: $\ket{1}-i\ket{-1}$}. The coincidence counts from the azimuthal scanning are presented in Fig.\ref{fig:fig3} (a), where interference fringes with a visibility of $0.92\pm0.01$ are observed, indicating that the path information has been erased, and equivalently, the OAM of the photon. The detection probability function is consistent with the theory of Eq.~(\ref{eq:prob}).  

Next, the polarizer angle was varied in the range $\alpha = 0$ through $\pi$ with subsequent measurements of the spatial pattern. The visibility of the interference fringes with respect to the polarizer orientation was calculated from the measured data and is presented in Fig.~\ref{fig:fig3} (b). The interference fringes are minimal at $\alpha=0$ and maximal when $\alpha=\pi/4$, as expected. Indeed, the polarizer controls the interference between the two OAM paths with a visibility proportional to $|\sin(2\alpha)|$, as predicted by the theory.

Finally, we performed a \BV{delayed measurement} variation of the quantum eraser by extending the path length of arm A by 2.3 m, with the experimental results presented in Fig.~\ref{fig:fig3} (c) and (d). The same procedures were used to mark the OAM paths and to erased the path information.  The visibility with respect to the variation of the polarizer angle was calculated and presented in Fig.~\ref{fig:fig3} (d), showing a range from $V=0.008\pm0.01$ to $V=0.96\pm0.02$, in good agreement with theory.

Complementarity between path information and fringe visibility is essential to the quantum eraser. By defining the two distinct paths using the OAM DoF, we have shown that through polarisation-OAM hybrid entanglement, it is possible to distinguish ($V=0.04\pm0.01$) and erase ($V=0.92\pm0.01$) the OAM path information of a photon through the polarisation control of its entangled twin. Our work is consistent with previous studies using entanglement and linear momentum of light \cite{walborn2002double}, as well as with angular fringes observed with weak classical light \cite{chen2014revisiting}, both of which used physical paths rather than abstract paths for the path interference.  Our \BV{delayed-choice} experiment highlights the extent to which information is made available to an observer through a \BV{delayed measurement} variation of the quantum eraser, where the analysis of the fringe pattern occurs before the decision to mark the paths (or not) is made. Indeed, we distinguished ($V=0.008\pm0.01$) and erased ($V=0.96\pm0.02$) the OAM path information, showing that causality does not play a role in the outcome path interference, which is a non-classical property of quantum mechanics. Furthermore, mutual exclusivity between the visibility and path information was demonstrated by varying the amount of OAM path information present in the system.
 
Significantly, our scheme shows the important role of hybrid entanglement which has been discussed previously as the main aspect of the quantum eraser \cite{neves2009hybrid,ma2013quantum}. Abstracting the path to OAM, with all the versatile tools that come with this choice of path, may provide the possibility of finding new approaches for studies in quantum information and communication.  While we note that in principle any degree of freedom may be used, OAM is an attractive choice due to the possibility to explore the impact of dimensionality in such systems, given that it offers an infinitely large Hilbert space in which to operate. Finally, our scheme contrasts previous reports that rely primarily on traditional path-phase interferometric methods, overcoming the sensitivity and complexity of such experiments. 
  
In conclusion, we have shown that the OAM of a photon may be treated as an abstract path, reminiscent of a slit. Using OAM-polarization hybrid entanglement, we have shown that, just as in the double slit quantum eraser, the OAM information of a photon that is marked with orthogonal polarizations can be erased through the polarization control of a bi-photon twin, both in the conventional scheme and in a \BV{delayed measurement} type arrangement. In both schemes the fringe visibility increases with a reduction in the OAM path information.
%%%%%%%%%%%%%%%%%%%%%%%%%%%%%%%%%%%%%%%%%%%%%%%%%%%%%%%%%%%%%%%%%%%%%%%%%%%%%%%%%%%%%%%%%%%%%%%%%%%%%%%%%%%%%%%%%%%%%%%%%%%%%%%%%%
%End of paper

\section{Acknowledgments}
The authors express their gratitude to Lorenzo Marrucci for providing q-plates and thank Melanie McLaren, Benjamin Perez-Garcia, Sandile Kumalo and Gareth Berry for their useful advice and insight. B.N. acknowledges financial support from the National Research Foundation of
South Africa and I. N. from the Department of Science
and Technology (South Africa).

%
%\bibliographystyle{apsrev4-1}
%\bibliography{mainref}

\end{document}